\documentclass[aps,prl,twocolumn,superscriptaddress,showpacs]{revtex4}
\usepackage{hyperref}

\usepackage{graphicx}

\usepackage{amsmath}
\usepackage{epsfig}

\begin{document}
\title{Experimental purification of coherent states}

\author{Ulrik L. Andersen}
\affiliation{Institut f\"{u}r Optik, Information und Photonik,  Max-Planck Forschungsgruppe, Universit\"{a}t Erlangen-N\"{u}rnberg, G\"{u}nther-Scharowsky str. 1, 91058, Erlangen, Germany}
\email{andersen@kerr.physik.uni-erlangen.de}
\author{Radim Filip}
\affiliation{Department of Optics, Research Center for Optics, Palacky University, 17. Listopadu 50, 77200 Olomouc
Czech Republic}
\author{Jarom\'{\i}r Fiur\' a\v sek}
\affiliation{Department of Optics, Research Center for Optics, Palacky University, 17. Listopadu 50, 77200 Olomouc
Czech Republic}
\author{Vincent Josse}
\affiliation{Institut f\"{u}r Optik, Information und Photonik,  Max-Planck Forschungsgruppe, Universit\"{a}t Erlangen-N\"{u}rnberg, G\"{u}nther-Scharowsky str. 1, 91058, Erlangen, Germany}
\author{Gerd Leuchs}
\affiliation{Institut f\"{u}r Optik, Information und Photonik,  Max-Planck Forschungsgruppe, Universit\"{a}t Erlangen-N\"{u}rnberg, G\"{u}nther-Scharowsky str. 1, 91058, Erlangen, Germany}

\date{\today}

\begin{abstract}
We propose a scheme for optimal Gaussian purification of coherent states from several imperfect copies. The proposal is experimentally demonstrated for the case of two copies of a coherent state sent through independent noisy channels. Our purification protocol relies on only linear optics and an ancilla vacuum state, rendering this approach an interesting alternative to the more complex protocols of entanglement distillation and quantum error correction.  
\end{abstract}


\maketitle

A central aim in quantum information science is to transfer, manipulate and store nonorthogonal quantum states with high fidelity. Unfortunately, however, any realistic processing of quantum states will inevitably be subject to environmentally induced noise, a process known as decoherence which transforms pure states into mixed states. 
For reliable processing of quantum states it is therefore of paramount importance to develop experimentally feasible methods that circumvent the effects of losses and decoherence. One of these methods is quantum error correction coding where the state under interrogation is protected by encoding it into a larger and more robust system~\cite{shor95.pra,steane96.prl}. Another scheme, which can be employed for faithful transmission of unknown quantum states, is perfect teleportation~\cite{bennett93.prl}. This can be accomplished by the use of maximally entangled states which have been obtained after entanglement distillation~\cite{bennett96.prl}. Finally, a very interesting alternative - which is the subject of this Letter - is to process several identically prepared quantum states, the number of which can be reduced to extract a state with higher purity after the states have been corrupted by the interaction with independent environments~\cite{cirac99.prl,keyl01.an,fischer01.pra}.  

Most attention has been focused on circumventing decoherence of states described by finite-dimensional systems such as the energy levels of an atom or the polarization of a photon. In this regime, experimental demonstration on quantum error correction coding~\cite{chiaverini04.nat} as well as entanglement distillation~\cite{kwiat01.nat,yamamoto03.nat,pan03.nat} enabling faithful transfer of quantum states via teleportation have been recently conducted. Furthermore, the purification protocol for qubits relying on an ensemble of input states was also recently experimentally demonstrated for the case where two qubits, which have been subject to the same noise, were used to derive a purer one~\cite{ricci04.prl}. 

In contrast, to date no experimental realizations on overcoming decoherence in infinitely dimensional systems, in which continuous variables (CV) carry the information~\cite{braunstein04.xxx}, have been performed. This is mainly due to the experimental difficulties associated with the physical implementation of the proposed schemes for CV quantum error correction~\cite{braunstein98.nat,lloyd98.prl}, which requires several highly squeezed beams, and the scheme of CV entanglement distillation requiring either non-Gaussian states or non-Gaussian operations~\cite{duan00.prl,browne03.pra,eisert02.prl}. 

In this Letter we propose and experimentally demonstrate a much simpler protocol undoing the effect of decoherence of continuous variable states. It is based on the noisy manipulation, storage or transmission of several copies of the state and then using the whole corrupted ensemble to extract an optimally purified state in the Gaussian regime. The proposed scheme is remarkably simple relying on only linear optics and an ancillary vacuum state. We will show that the added error variance of the purified state will be reduced by a factor equal to $1/N$ where $N$ is the number of prepared input states. The method is an extension of the purification protocol for qubits mentioned above into the CV regime. However, it is worth mentioning that in the qubit based purification protocol the scheme is probabilistic whereas our protocol is unconditional, an important advantage in the continuous variable domain.

Let us first illustrate the effect in the simplest case of two coherent states $|\alpha\rangle^{\otimes 2}$, denoted by 1 and 2 and displaced according to the information they carry. We characterize these states by the annihilation operators $\hat a_{1,2}=1/2(\hat x_{1,2}+i\hat p_{1,2})$, where $\hat x$ and $\hat p$ are the pair of non-commuting Hermitian operators associated with the amplitude and phase quadratures of the electro-magnetic field ($[\hat x,\hat p]=2i$). These two states are then sent through noisy Gaussian channels or into some noisy Gaussian operations, from which they emerge as Gaussian mixed states, $\rho_m^{\otimes 2}$. In the Heisenberg picture, the quadrature operators of the mixed states can be expressed as 
\begin{eqnarray}
\hat{x}_{m1,2}=\hat{x}_{in1,2}+\hat{x}_{N1,2}\nonumber\\
\hat{p}_{m1,2}=\hat{p}_{in1,2}+\hat{p}_{N1,2},
\label{noise}
\end{eqnarray}    
where $\hat{x}_{N1,2}$ are noise operators representing the additive Gaussian noise of the imperfect quantum channel or operation. This noise might originate for instance from spontaneous emission in a linear amplifier which is incorporated in a transmission line to compensate for losses. We assume the noise to be uncorrelated (which indeed is the case for the linear amplifier) and hence described by a covariance matrix with only the diagonal elements being non-zero: $Cov=diag(\Delta ^2\hat x_{N1},\Delta^2\hat p_{N1},\Delta^2\hat x_{N2},\Delta^2\hat p_{N2})$, where $\Delta^2\hat{q}=\langle\hat{q}^2\rangle-
\langle\hat{q}\rangle^2$ denotes the variance of $\hat{q}$. Furthermore, since the two input states are identically prepared we have $\langle \hat x_{in1}\rangle=\langle\hat x_{in2}\rangle$ and $\langle\hat p_{in1}\rangle=\langle\hat p_{in2}\rangle$.

Having only a single state at hand, the linear transformation in Eq. (\ref{noise}) cannot be reversed. However having two copies at our disposal one can purify the state or in other words one can reduce part of the excess noise while keeping the expectation value constant. The purification procedure runs as follows. In a first step the two corrupted copies are mixed in-phase on a 50/50 beam splitter concentrating the information into a single spatial mode, and in a second step this mode is combined with vacuum on another 50/50 beam splitter, from which the outputs appear as purified states. The transformation is represented by
\begin{eqnarray}
\hat{x}_{pu}=\frac{1}{2}(\hat{x}_{in1}+\hat{x}_{in2}+\hat{x}_{N1}+\hat{x}_{N2})+\frac{1}{\sqrt{2}}\hat{x}_{v}\\
\hat{p}_{pu}=\frac{1}{2}(\hat{p}_{in1}+\hat{p}_{in2}+\hat{p}_{N1}+\hat{p}_{N2})+\frac{1}{\sqrt{2}}\hat{p}_{v}
\end{eqnarray}    
where $\hat{x}_v$ and $\hat{p}_v$ are the noise operators for the vacuum state. We clearly see that the purification scheme is universal (mean value preserving) and covariant (phase insensitive) as required by a purification protocol. Assuming the added thermal noise to be symmetric (corresponding to equal diagonal elements in the covariance matrix) and equal to $\lambda$, the error variance of the purified state is
\begin{eqnarray}
\sigma^2=1+\frac{\lambda}{2}
\label{variance}
\end{eqnarray}    
This should be compared with the error variance of the states before purification which is $\sigma^2=1+\lambda$, thus the protocol reduces the added noise variance by a factor of 2. We also note that, in fact, two purified copies are produced. However, since these copies are classically correlated (and hence dependent), they cannot be used in harmony for further processing. Therefore in order to finalize the purification scheme, one of the copies must be discarded. Let us finally note that the idea of using quantum interference for purification was also used in the error filtration protocol of ref. \cite{lamoureux05.prl} to purify noisy qubits.

To characterize the quality of our purification protocol we used the fidelity $F=\langle\alpha|\rho|\alpha\rangle$ which quantifies the overlap between the input coherent state, $|\alpha\rangle$, and the purified state, $\rho$. For Gaussian states the fidelity can simply be read as $F=2/\sqrt{(1+\Delta^2\hat x)(1+\Delta^2\hat p)}$
if the expectation values are conserved (corresponding to unity gain operation). For the purification protocol outlined above the gain is unity and the fidelity becomes 
\begin{equation}
F=\frac{1}{1+\frac{\lambda}{4}}
\label{fidelity}
\end{equation}    
which should be compared with the fidelity for the non-purified states: $F=1/(1+\frac{\lambda}{2})$.    

Let us show that the performance of our quantum purification protocol beats that of the standard classical purification protocol usually employed when one deals with classical information. The classical strategy is based on an optimal measurement followed by re-preparation of the state. For coherent states with added thermal Gaussian noise the optimal measurement strategy is to simultaneously measure the amplitude and the phase quadrature. This can be accomplished by splitting the state on a 50/50 beam splitter and subsequently measure two conjugate variables, one in each output. In the two copy case, the purified state can be prepared based on two estimates and has an error variance of $\sigma^2=2+\lambda/2$, which is clearly larger than the error variance in Eq. (\ref{variance}) associated with the quantum approach.

\begin{figure}[h] \centering \includegraphics[width=6cm]{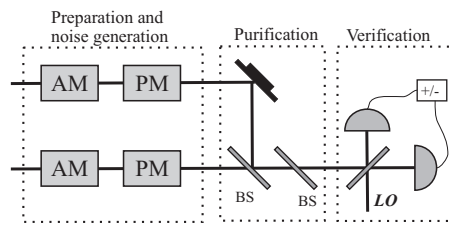} \caption{\it Schematic diagram showing the experimental setup for coherent state purification. The setup is divided into three boxes which define the preparation and noise generating stage, the purification stage and the verification stage. AM: Amplitude modulator, PM: phase modulator, BS: 50/50 beam splitter and LO: Local oscillator.}
\label{fig1}  \end{figure}

In the experimental implementation of the proposed scheme we define our coherent quantum states to reside at frequency sidebands within a well-defined bandwidth of an electro-magnetic field. This ensures a truly pure coherent state and allows easy control of the coherent amplitudes via simple electro-optic modulators operating at the sideband frequency. The laser used for the experiment is a monolithic ND:YAG laser producing a field at 1064~nm to generate the two copies of coherent states by displacing the vacuum sidebands using two amplitude modulators and two phase modulators and to provide a local oscillator for verification of the protocol.

Decoherence of the coherent state corresponds to phase space displacements of the sidebands with uncorrelated Gaussian noise. This noise was actively applied to the states using the same four modulators as were used for the coherent state preparation. In order to avoid any correlation between $\hat{x}_1$, $\hat{x}_2$, $\hat{p}_1$ and $\hat{p}_2$ we used four independent noise generators and the modulators were carefully aligned to ensure a high purity of the modulations. The variances of the noise added to all four quadratures were equalized hereby ensuring that the decoherence effect is phase-covariant. 

In the purification step the two copies are carefully mode-matched (with a visibility of $97.0\pm 0.2\%$) on a 50/50 beam splitter and actively locked to constructively interfere using an electronic feedback loop. Subsequently the purified state is divided by another 50/50 beam splitter.  
 
We used a single homodyne detector to characterize the properties of the pure input coherent states, the corrupted states and the purified state. However, when analyzing the input coherent states as well as the decohered states we corrected the results taking into account the degradation induced by the two 50/50 beam splitters. Furthermore, precise determination of all error variances was ensured by inferring out the overall homodyne detector efficiency (measured to $\eta=80\pm 1\%$) from the measurement results.    

To confirm that there is no intra correlation of the noise between the amplitude and phase quadrature we scanned the local oscillator when monitoring the individual mixed states. We observed no variation of noise power as a function of the phase of the local oscillator, which is a clear signature of uncorrelation between the quadratures.  

\begin{figure}[h] \centering \includegraphics[width=7.5cm]{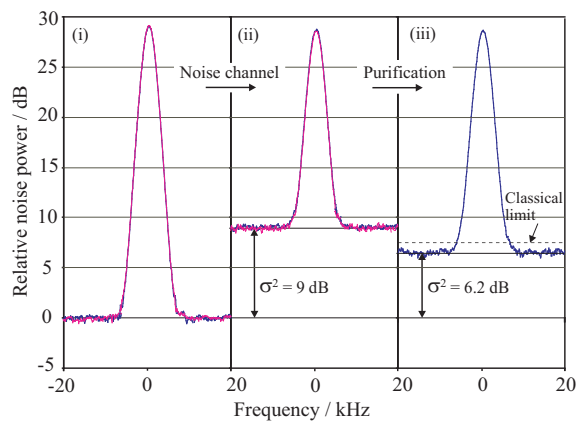} \caption{\it Power spectra showing the purification protocol. (i) Power spectra for two identical but independently prepared coherent states with an excitation corresponding to about 736 photons. (ii) Power spectra for the two states after being subjected to environmental induced Gaussian noise. (iii) Power spectrum for the purified copy. All spectra are normalized to the quantum noise level and centered around 15MHz. The resolution bandwidth is 3kHz, the video bandwidth is 30Hz and we plot an average of 10 traces.}
\label{fig2}  \end{figure}

In Fig. \ref{fig2} we show the results for a certain purification run where symmetric Gaussian noise with variance $\lambda=7$ has been added to the pure states. Here we plot the power spectra in the frequency range from 14.98MHz to 15.02 MHz for phase quadrature of the input coherent states (i), the corrupted states (ii) and the purified state (iii). While the size of the modulation depths stay constant (corresponding to unity gain), the noise floors vary in level.   
We clearly see the effect of purification, the uncertainty of the purified state being $\sigma^2=4.2\pm0.3$ in contrast to $\sigma^2=8\pm0.8$ before the purification. With unity gain this corresponds to an increase in fidelity from $F=0.21\pm0.02$ to $F=0.38\pm0.01$. We also note that the corresponding expected fidelity obtained using the classical approach is $F=0.31$, a value which is clearly surpassed by our quantum approach. 

As a result of the non-perfect mode-matching efficiency in the spatial coupling between the two input modes, deviations from unity gain were encountered in the experiment. The gains, defined by $g_{x}= \langle x_{pu}\rangle/\langle x_{m}\rangle$ and $g_{p}=\langle p_{pu}\rangle/\langle p_{m}\rangle$, were determined to be $g_{x}=0.96\pm0.01$ and 
$g_{p}=0.99\pm0.01$. As a result of this deviation from unity gains, the fidelity becomes dependent on the complex amplitude of the input coherent state. Therefore the figure of merit is defined as an average of the "single-shot" fidelities: 
$F_{ave}=\int F(\alpha) P(\alpha) d^2\alpha$ where $P(\alpha)$ is the probability for drawing a certain coherent state, $|\alpha\rangle$.
E.g. considering a Gaussian distributed set of input coherent states with an uncertainty in photon number of $100$ 
we find $F_{ave}=0.37$ for the example in Fig.~\ref{fig2}. Thus, despite the non-unity gain operation only a small decrease in the fidelity is encountered if a restricted set of input states is considered.

\begin{figure}[h] \centering \includegraphics[width=7.5cm]{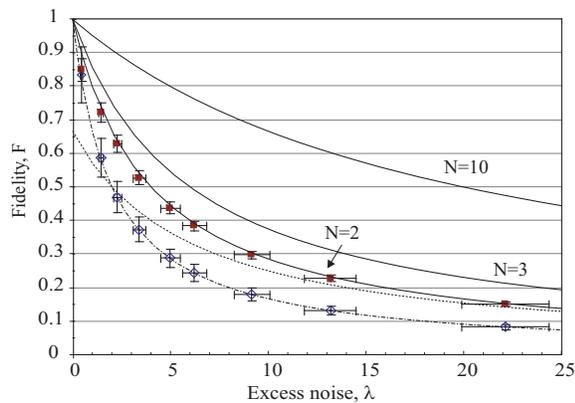} \caption{\it Fidelity as a function of the excess noise introduced in the noisy environment. Here the open diamonds are the experimental results before purification while the filled squares are results after purification. For comparison we included the theoretical predictions for the fidelities before purification (dot-dashed line), after purification (lower solid line) and the classical limit (dotted line) for the two-copy case. The two upper solid lines represent the purification protocol for 3 and 10 copies.}
\label{fig3}  \end{figure}

We demonstrate the purification effect for different amounts of added noise in  Fig.~\ref{fig3} where we plot the fidelity as a function of the excess noise, both before and after purification of the decohered state. In these measurements the noise variances were investigated precisely at 15~MHz (with a bandwidth of 100~kHz) by turning off the modulations and monitoring the noise power over 2 seconds. We then calculate the fidelity using Eq. (\ref{fidelity}). The dot-dashed line in Fig.~\ref{fig3} represents the fidelity after propagation in a noisy environment, whereas the solid lines are the theoretical predicted fidelities after purification by the use of 2, 3 or 10 copies (see theory below). Nice agreement is observed between the experimental values and the theoretical curve for $N=2$. Finally, for comparison we have also included the expected fidelities for classical purification for $N=2$, represented by the dotted curve.

Let us now consider the general problem where $N$ identical coherent states are subject to a noisy transmission and subsequently  purified. First, we concentrate $N$ noisy replicas into a single mode using a network of beam splitters resulting in a noisy state of amplitude $\sqrt{N}\alpha$. Secondly, the state is purified by the use of a single beam splitter and the overall Gaussian transformation can be described by the transformation
\begin{equation}
q'=a\sum_{j=1}^{N}q_j+bq_{anc}
\end{equation}  
where $q'$ is an arbitrary quadrature amplitude of the output mode, $q_j$ are the corresponding quadratures of the modes input to the protocol and $q_{anc}$ is an ancilla mode. Universality of the transformation, i.e. conservation of the mean values, is guaranteed by choosing $a=1/N$ and $\langle q_{anc}\rangle=0$. By means of the commutation relations for every conjugate pair of quadrature amplitudes we find $b=\sqrt{\frac{N-1}{N}}$. The variance of the purified state is then $\sigma^2=\frac{1}{N^2}\sum_{j=1}^N \Delta^2\hat q_j+\frac{N-1}{N}\Delta^2\hat q_{anc}$. In order to ensure phase-insensitive operation, the ancilla state is vacuum and the error variance reduces to
\begin{equation}
\sigma^2=1+\frac{\lambda}{N}
\end{equation} 
which coincides with Eq. (\ref{variance}) for N=2. The purification fidelity for N copies is $F=1/(1+\frac{\lambda}{2N})$, which clearly approaches 1 as the number of copies increases. This fidelity is plotted in Fig.~\ref{fig3} for N=2,3 and 10. 
The experimentally feasible setup for the N-copy purification protocol comprises N laser beams, a standard beam splitter array (with N-1 beam splitters) followed by a single beam splitter. Such a network can be compactly built using integrated optics. Alternatively, all copies can be carried by a single laser beam: The beam is divided into N time bins where each time bin is associated with a copy of the state. By defining the states in this way, the copies can then be efficiently mixed into a single time bin (or state) by using an optical loop with a fast switch to control the mixing ratios. 
The details of such a scheme will be elaborated in another article~\cite{filip05}. 

Let us stress that an analogous scheme also works when the information is carried by squeezed states rather than 
coherent states. In this case, the information is encoded as a displacement of the squeezed
quadrature and the ancillary mode is required to be vacuum squeezed with a degree and an angle equal to the input modes in order to approach optimal purification.
 
In conclusion, in this Letter we have proposed and experimentally demonstrated an easily implementable scheme for purifying coherent states which have been corrupted by a noisy operation or transmission. It is based on the use of an ensemble of identically prepared pure states, which have been under the influence of independent noisy environments. Then by concentrating the noisy outputs into a single spatial mode it is possible to optimally purify the state by the use of a beam splitter with appropriate beam splitting ratio. We stress that our scheme does not, as opposed to previous proposals on noise-free transmission, rely on experimentally challenging entanglement distillation protocols which requires non-Gaussian operations\cite{duan00.prl}. 
In our experiment we obtained clear evidence for coherent state purification in the two-copy case and showed that the classical limit was surpassed. This is the first experimental demonstration of quantum state purification in the continuous variable regime, a result which is likely to play an important role in combating decoherence in the operation of future quantum gates, memories and transmission channels. 

We thank Ladislav Mi\v sta Jr. and Petr Marek for fruitful discussions.
This work has been supported by the EU project COVAQIAL (project no. FP6-511004), DFG (the Schwerpunkt programm 1078), the network of competence QIP (A8), the project 202/03/D239 of Grant Agency of Czech Republic and the grant MSM6198959213 of the Ministry of Education of Czech Republic. ULA acknowledges an Alexander von Humboldt fellowship.

\end{document}